# Convolutional Neural Networks for Predictive Modeling of Lung Disease

Yingbin Liang *
Northeastern University
Seattle, USA

Xiqing Liu
Columbia University
New York, USA

Haohao Xia
University of Southern California
Los Angeles, USA

Yiru Cang
Northeastern University
Boston, USA

Zitao Zheng
Independent Researcher
New Jersey, USA

Yuanfang Yang
Southern Methodist University
Dallas, USA

*Abstract*: In this paper, Pro-HRnet-CNN, an innovative model combining HRNet and void-convolution techniques, is proposed for disease prediction under lung imaging. Through the experimental comparison on the authoritative LIDC-IDRI dataset, we found that compared with the traditional ResNet-50, Pro-HRnet-CNN showed better performance in the feature extraction and recognition of small-size nodules, significantly improving the detection accuracy. Particularly within the domain of detecting smaller targets, the model has exhibited a remarkable enhancement in accuracy, thereby pioneering an innovative avenue for the early identification and prognostication of pulmonary conditions.

## I. INTRODUCTION

In the field of medicine, precision medicine and personalized treatment are gradually becoming the mainstream trend, and the early diagnosis and prediction of lung diseases is a key link in this trend. Respiratory ailments, encompassing conditions like carcinoma of the lung, chronic obstructive pulmonary disease (COPD), and pneumonic infections, not only boast a considerable prevalence rate but also profoundly influence the standard of living and longevity of those afflicted [1]. Traditional diagnostic methods rely on the experience of doctors and various imaging examinations, such as X-rays, CT scans, etc., but in complex cases, these methods often have certain limitations, such as the subjectivity of diagnosis, sensitivity, and lack of specificity.

Over recent years, the swift advancement in artificial intelligence, notably the leaps in deep learning, has ushered in novel prospects for the prognosis and predictive analysis of lung diseases [2]. Serving as potent tools for pattern recognition, deep neural networks are capable of autonomously extracting features from voluminous datasets, devoid of manual intervention[3-5]. This self-learning capability renders them exceptionally promising in tackling intricate image recognition challenges, particularly in medical diagnostics[6]. Especially in lung image analysis, deep neural networks can identify subtle structural changes that are difficult to detect with traditional methods, thus improving the accuracy of disease detection[7]. However, developing an efficient and precise lung disease prediction model leveraging deep neural networks is far from straightforward. First, the construction of high-quality lung image data sets is the foundation, which requires a sufficient number and diversity of samples, as well as accurate labeling information. Secondly, the design of the model should fully consider the characteristics of lung imaging to ensure the effectiveness and robustness of the model. Ultimately, the model's training and validation phases necessitate robust methodological underpinning to ascertain its generalizability and predictive precision.

The study aims to develop a robust and precise model for predicting lung diseases by leveraging deep neural network technology and analyzing lung image characteristics. This research will encompass the creation of the dataset, the design of the model, optimization techniques during training, and the verification of results. The ultimate goal is to innovate new methods for the early diagnosis and prediction of lung diseases. By doing so, we seek to enhance the accuracy of clinical decision support, provide earlier treatment opportunities for patients, and ultimately advance the entire healthcare system towards a future marked by greater precision and efficiency.

## II. CORRELATIONAL RESEARCH

Recently, the integration of artificial intelligence with image analysis techniques has significantly advanced the field of intelligent diagnostic capabilities in medical imaging, leading to substantial technological breakthroughs[8-10]. Researchers globally are dedicated to enhancing the performance of intelligent diagnostic systems, creating computer-aided diagnostic tools that combine machine learning and deep learning to improve the recognition and classification of lung images[11]. Despite these advancements, the inherent complexity and morphological variations in lung imaging continue to pose significant challenges. These challenges result in notable deficiencies in the sensitivity and classification accuracy of current detection technologies, which must be overcome to achieve optimal performance[12].

Lassen et al. [13] used regional growth techniques to identify pulmonary nodules, and the initial seed points needed to be manually determined. However, the varied characteristics of pulmonary nodules and their resemblance to surrounding tissue pose significant challenges, constraining the precision of this approach. Farahani FV et al. [14] integrated multi-layer perceptrons, K-nearest neighbors and SVM classifiers, and determined the categories of pulmonary nodules by voting mechanism through CT image segmentation and shape feature analysis. Javaid et al. [15] used the intensity threshold and K-means algorithm to first segment and then refine to realize the preliminary detection of pulmonary nodules. Roth et al. [16] used 2D CNN to detect lung nodules and reduce misjudgment through image preprocessing and feature extraction. In recent years, the emergence of more efficient 2D CNN models has promoted the development of related fields.

Yan et al. [17] proposed an innovative cancer detection scheme that integrates graph convolutional neural networks (GCNs) and advanced image analysis directly applied to lung cancer detection. Similar to how GCNs enhance prognosis accuracy in gastric and colon cancers by analyzing spatial relationships in tumors, this method could improve the detection and predictive modeling of lung cancer. Adapting this technique to lung cancer could enhance tumor detection and staging, ultimately improving treatment strategies and patient outcomes. This highlights the potential of convolutional neural networks in advancing lung disease management. Setio et al. [18] focused on the multi-angle analysis of CT images. They used two-dimensional slices of lung nodules in different directions (including axial, coronal, sagittal and six diagonal angles) to build a multi-view 2D CNN model to capture the spatial distribution characteristics of lung nodules and further optimize the classification effect. Liu et al. [19] presented a comprehensive study on feature extraction using CNNs, emphasizing the model's capability to identify intricate patterns in medical images. Lin et al. [20] focused on constructing disease prediction models using advanced deep learning technologies, demonstrating how AI can be leveraged to predict medical conditions by analyzing complex datasets. Hu et al. [21] investigated multi-scale image fusion systems in intelligent medical image analysis, enhancing image analysis accuracy through multi-scale techniques, a concept that parallels the void-convolution technique used in our Pro-HRnet-CNN model. Sun et al. [22] explored the optimization of Natural Language Processing (NLP) models using multimodal deep learning approaches, with principles of multimodal learning and model optimization pertinent to improving the performance of CNNs in medical image analysis. Xiao et al. [23] examined the use of attention mechanisms in deep learning models for mining medical textual data, providing valuable insights into improving model accuracy and efficiency. Additionally, Skourt et al. [24] employed the U-net framework to analyze CT images, attaining precise delineation of lung nodules. The model combined downsampling encoding and upsampling decoding to extract and recover features effectively, and obtained a Dyess similarity coefficient of 0.9502 on the LIDC-IDRI dataset.

These studies collectively contribute to the foundational knowledge required to develop sophisticated deep learning models for medical diagnostics. Our proposed Pro-HRnet-CNN model builds on these advancements, integrating HRNet and void-convolution techniques to enhance the detection accuracy of small pulmonary nodules, thereby pushing the boundaries of early lung disease diagnosis and prediction.

## III. METHOD

### A.   Deep neural network

#### 1. Void convolution

The core components of convolutional neural network include convolutional layer, activation unit, downsampling module and fully connected layer. The convolution layer realizes the high-level abstraction of image features by performing convolution operations. The downsampling function can be realized by convolutional processing or pooling technology with step size greater than 1. Studies have confirmed that the use of large-step volume instead of traditional pooling can significantly optimize the overall performance of the network[25-27]. However, with the image undergoing multiple downsampling, the original details tend to be reduced. In this case, the hollow convolution technique can partially alleviate this problem and maintain the feature integrity. In void convolution, as shown in formula (1), the void rate l is incorporated into the properties of the convolution filter, and its function is to intersperse voids on the basis of standard convolution, so as to expand the perceptual field of view of the feature map. A detailed void-convolution structure is shown in Figure 1.

$$(F *_l k)(p) = \sum_{s+lt=p} F(s)k(t) \quad (1)$$

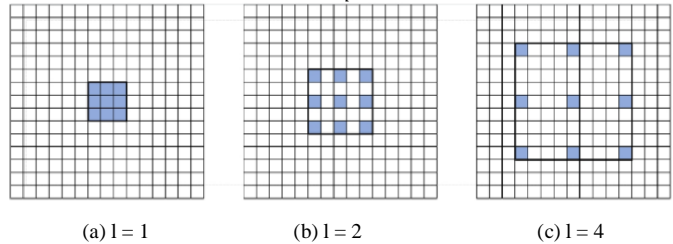

(a) l = 1    (b) l = 2    (c) l = 4

Figure 1. Convolution kernel with different void rates

In Figure 1, the blue boxes indicate three types of 3x3 convolution nuclei with different cavity rates: standard (a), nuclei with cavity rate l=2 (b), and more advanced (c). From the 3x3 perception domain (a) to 7x7 (b) to 15x15 (c), the void convolution amplifies the perception domain step by step, far exceeding the 7x7 limit of three successive standard 3x3 convolution layers. This strategy can greatly expand the perception field without increasing parameters and reduce the information loss of small targets while maintaining high resolution.

#### 2. High resolution deep neural networks

In the field of image analysis, especially in the field of small scale target recognition, it is very important to maintain high definition feature representation. Based on this need, the researchers designed a deep learning framework specifically optimized for high-resolution performance, High resolution Neural networks (HRNet), which is particularly suitable for complex visual tasks such as human pose resolution and scene semantic understanding. As shown in Figure 2, HRNet's innovative architecture ensures efficient flow of information at multiple scales, resulting in superior performance in small target detection.

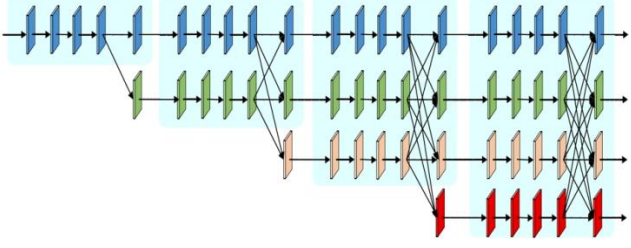

Figure 2. HRNet architecture diagram

HRNet's architecture is subdivided into four progressive stages, the fourth of which is visualized as four layers of rectangular blocks arranged from top to bottom, each layer distinguished by different colors, symbolizing the descending resolution of the feature map. Different from the linear flow of the traditional series feature extraction network, HRNet is unique in parallel with multiple resolution feature maps throughout the whole process, ensuring that the network always retains high definition features, which cleverly avoids the potential loss of small target information in the hierarchical transmission of feature maps. Furthermore, HRNet establishes an information exchange mechanism between high and low resolution feature maps, which significantly enhances the feature description power of the model. Therefore, in the face of the small target challenge in object detection, HRNet presents a clear advantage with its excellent feature capture capability.

3. Evaluation index

In this study, we adopted Average Precision (AP), a commonly used evaluation criterion in object detection and instance segmentation, to measure model performance. In the context of target detection, IOU serves as a prevalent quantitative metric, adeptly assessing the overlap between the predicted bounding box and the actual target area, thereby providing an efficacious gauge for the precision of object detection algorithms, and thus reflect the accuracy of detection results. The specific calculation formula of the intersection ratio is shown in expression (2).

$$IoU = \frac{Area_{PRED} \cap Area_{GT}}{Area_{PRED} \cup Area_{GT}} \quad (2)$$

The Intersection over Union (IoU) metric is calculated by determining the ratio of the intersecting area between the predicted bounding box (the detection frame) and the true bounding box (the actual frame), to the combined area encompassed by both boxes when they are merged together. The higher the value, the greater the overlap degree of the two frames, the better the detection performance. In the average accuracy (AP) evaluation process, an IoU threshold is usually set in advance to judge whether the detection frame classification is correct or not.

By iterating over all the bounding boxes predicted by the model, we can calculate Precision and Recall based on the given equation, such as formula (3) and formula (4).

$$Precision = \frac{TP}{TP + FP} \quad (3)$$

$$Recall = \frac{TP}{TP + FN} \quad (4)$$

The accuracy rate reflects the proportion of actual positive cases in the current predicted detection box; The recall rate measures the proportion of correctly detected boxes to all actual positive boxes. Each time a detection box is evaluated, a set of corresponding Precision and Recall values is generated. Plot these pairs of values to form an accuracy rate-recall curve, or P-R curve. According to this curve, we can calculate the average accuracy of the model according to formula (5).

$$AP = \int_0^1 Precision(Recall)\,d(Recall) \quad (5)$$

The AP is the area under the P-R curve. Each category is calculated independently, and the AP average of all categories is the mAP. Because there is only one kind of label in the data set, AP is directly used as the index. The AP calculation in instance segmentation only needs to change the detection frame and true frame to prediction and true mask respectively, and the other steps are the same as the target detection.

4. Loss function

Conventional GAN architectures employ stochastic noise as input to produce outputs, without considering the correlation between the quality of the input imagery and the resultant output labels. In the field of image segmentation, the generative adversarial network generates segmentation labels by parsing the input image, ensuring that the output directly responds to the input pixels, forming a clear function mapping. The objective function for the complete task is specified in Equation (6).

$$L(p,t) = \lambda L_{cls}(p, p^*) + p^* L_{reg}(t_i, t_i^*) \quad (6)$$

Here λ is used to coordinate the classification and regression losses and takes a value of 0.5. Where, p refers to the prediction probability, and p* is the truth label of the Anchor box. An anchor box is deemed a positive sample, denoted as p* = 1, if its Intersection over Union (IoU) with the actual boundary exceeds 0.5. Conversely, an IoU less than 0.2 classifies it as a negative sample, represented as p* =0.

In the classification task, we choose weighted binary cross entropy as the loss function, and its specific mathematical expression is shown in formula (7). Here, the symbol w represents the weight factor used to balance the different classes of loss contributions.

$$L_{cls}(p, p^*) = -w[p^* \log p + (1 - p^*)\log(1 - p)] \quad (7)$$

The loss of position regression is calculated using the smooth L1 loss function specified in equation (8) to accurately quantify the position deviation between the predicted box and the real box.

$$L_{reg}(t_i, t_i^*) = \sum_{i \in \{x,y,z,d\}} smoot_{L1}(t_i - t_i^*) \quad (8)$$

$$smoot_{L1}(t_i - t_i^*) = \begin{cases} 0.5(t_i - t_i^*)^2, & if\,|t_i - t_i^*| < 1 \\ |t_i - t_i^*| - 0.5, & else \end{cases} \quad (9)$$

B. *Improved Pro-HRnet-CNN model*

This paper innovatively proposes the Pro-HRnet-CNN network structure, as shown in Figure 3. In the HRNet architecture, a rich feature pyramid is built by fusing feature maps at different levels. This design ensures that the final output features not only contain fine high-resolution details, but also incorporate rich semantic information extracted by the deep network, which is particularly important when dealing with fine structures such as small target nodules. Next, the candidate region (anchor box) generated by the RPN module is obtained by convolution operation, and then the class probability and position adjustment parameters of each anchor box are calculated by the two-layer fully connected network. After NMS (non-maximum suppression) screening, the candidate boxes that perform well are fed into the RoI Align module for precise cropping. Then, the multilevel cascade detection mechanism uses a series of detectors with increasing IoU thresholds to perform detailed classification and positioning correction of the candidate regions, and finally maps the detection results back to the original image space to achieve accurate positioning. This improved Pro-HRnet-CNN network structure not only ensures the capture of deep semantic information, but also effectively retains the key position clues of small-scale targets, and significantly improves the detection performance of small target nodules.

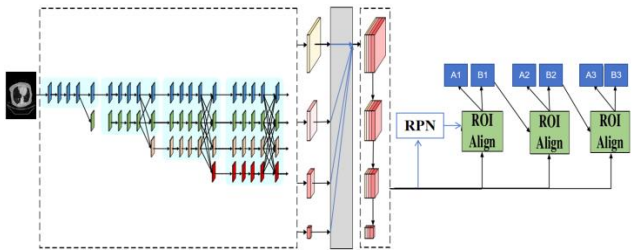

Figure 3. Improved Pro-HRnet-CNN model structure

IV. APPLICATION OF MODEL

A. Dataset

This study selected as the experimental dataset the LIDC-IDRI dataset, a lung CT image database supported by the National Cancer Institute of the United States, containing more than 1000 chest CT scans, each of which was labeled by four physicians with lung nodule boundaries, malignancy, and type. This dataset covers a diverse set of cases and is suitable for the development and testing of lung nodule detection algorithms. Open to researchers for non-commercial research, it is a key resource for lung disease imaging analysis and contributes significantly to improving lung nodule recognition technology.

In the experimental design, we adopted a dataset consisting of 6002 chest CT images, among which 4733 images were selected for constructing the training set through random sampling, and the remaining 1269 images were assigned to the test set, ensuring that the ratio of training to test data was close to 4:1, and there was no crossover between the two sets. All images are derived from the DICOM compliant LIDC-IDRI database. At the beginning of the experiment, we used the inherent Rescale Slope and Intercept attributes in DICOM file to convert the original pixel value of CT image into CT value (in HU) according to formula (10), and realized effective conversion and standardized processing of image value.

$$OutputUnits = m*SV + b \qquad (10)$$

In formula (10), variable "SV" represents the stored value of each pixel in the medical image in DICOM format, "m" corresponds to the Rescale Slope attribute, and "b" is the value of the Intercept attribute. After calculation, "OutputUnits" reflects the corresponding HU values of each position in the converted CT image. This transformation ensures consistency of image data and accuracy of clinical interpretation.

B. Experimental environment and parameters

1. Experimental environment

(1) The hardware platform used in the experiment is a high-performance server equipped with Intel(R) Core(TM) i9-9900X CPU with a main frequency of 3.50GHz, supplemented by 64GB memory and equipped with dual NVIDIA GeForce GTX 2080Ti Gpus. Each GPU has 11GB of video memory, ensuring efficient computing power for complex models.

(2) In terms of software environment, the operating system is a stable version of Ubuntu 16.04 LTS, The integrated development environment (IDE) selected is PyCharm, with Python 3.6 serving as the coding language. For the deep learning framework, both TensorFlow and PyTorch are employed to facilitate swift algorithm refinement and enhancement. Such hardware and software configuration provides powerful technical support for the experiment, ensuring the smooth progress of the experiment and the reliability of the results.

2. Experimental parameters

Considering the limitations of GPU memory, we split the complete CT image into 96x96x96 three-dimensional cube fragments to better fit the data processing capacity of the GPU, and set the Batch size of each batch to 8. In order to avoid overfitting of the model, data enhancement strategies were implemented for 3D image fragments (positive samples) containing nodules: first, random scaling according to [0.75,1.25] intervals; The second is to perform horizontal flip and random Angle rotation. The model optimization was carried out using the Stochastic Gradient Descent (SGD) technique, with the overall training cycle being fixed at 30 epochs.

Throughout the training procedure, we noted that with fewer than 30 epochs, despite the rapid decline in the loss values for both the training and validation datasets, there remains potential for enhanced model efficacy. After 30 epochs, loss reduction slowed down, additional training time cost increased, but performance improvement was limited. Based on this, we decided to stop training at 30 epochs.

During testing, we streamlined the process by initially discarding nodules under 3mm, then applying Non-Maximum Suppression (NMS) to manage overlapping candidate regions, simplifying the final results. To mitigate the risk of overfitting, our methodology encompassed augmenting the data, executing 10-fold cross-validation, and instituting a dropout protocol with a 0.5 dropout rate throughout training. These interventions were collectively geared toward fortifying the model's generalization prowess.

C. Experimental results and analysis

In order to gauge the capability of the newly proposed Pro-HRnet-CNN model in this chapter for accurately detecting small pulmonary nodules in medical images, we used AP under a specific IoU threshold as the main performance measure. At the same time, in order to make a comprehensive comparison, the

RetinaNet model widely recognized by the industry and the Fully Convolutional One-Stage Object Detection (FCOS) model are placed in the same data set environment for control tests, the precise comparative outcomes are elaborated upon in Table 1. In this way, the difference of detection accuracy and efficiency of each model is visually demonstrated.

TABLE I. COMPARISON OF TEST RESULTS OF DIFFERENT METHODS (%)

| Models | Basic algorithm | Backbone algorithm | AP | AP0.5 | AP0.75 |
|---|---|---|---|---|---|
| 1 | RetinaNet | ResNet-50 | 32.58 | 66.28 | 26.25 |
| 2 | FCOS | ResNet-50 | 35.27 | 74.05 | 28.19 |
| 3 | Pro-HRnet-CNN | HRnet | 44.16 | 79.93 | 43.33 |

According to the data in Table 1, compared with the RetinaNet and FCOS models, the average accuracy (AP) of the Pro-HRnet-CNN proposed in this chapter in the pulmonary image nodoid recognition task is significantly improved by 11.58% and 8.89%, respectively. Through careful comparison and in-depth analysis of the experimental results of various detection schemes, we confirmed that the Pro-HRnet-CNN architecture showed better applicability and higher accuracy in the detection of small and medium-sized lung nodules in medical images, surpassing other common models in the field of target detection at present.

In addition, we carefully planned a series of ablation studies to systematically evaluate the practical utility and contribution of the detection algorithms selected in this chapter and their combination with various backbone networks in the pulmonary medical image analysis task. These experiments not only validate the importance of individual components, but also demonstrate the significant effect of the overall architecture in improving detection accuracy. The comparison is shown in Table 2.

TABLE II. COMPARISON OF MODEL DETECTION RESULTS OF DIFFERENT BACKBONE NETWORKS (%)

| Models | Backbone algorithm | AP | AP0.5 | AP0.75 |
|---|---|---|---|---|
| Faster R-CNN | ResNet-50 | 38.78 | 76.54 | 36.13 |
| Pro-HRnet-CNN | ResNet-50 | 43.21 | 78.97 | 38.36 |
| Pro-HRnet-CNN | HRnet | 45.56 | 81.64 | 43.07 |
| Pro-HRnet-CNN | Pro-HRnet | 47.12 | 82.95 | 45.52 |

Based on the analytical insights from Table 2, it's evident that there exist significant disparities in the operational effectiveness of various detection models when tasked with lung nodule identification. The selection of a feature extraction network bears a direct influence on the model's precision in detection. Contrasting ResNet-50, opting for HRNet as the foundational network markedly amplifies the model's proficiency in discerning smaller nodule instances. Especially, when the IOU threshold is set to 0.75, the average accuracy (AP) is improved by 9.39%, which highlights the advantage of HRNet in capturing detailed features.

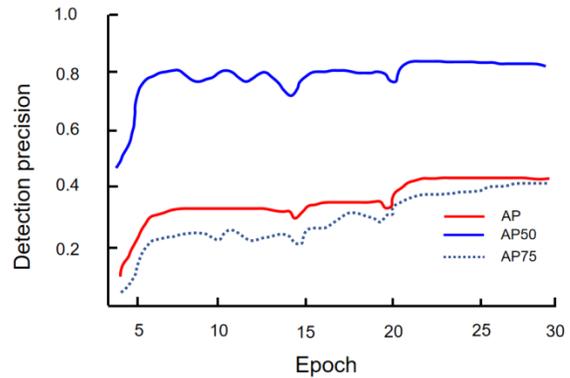

Figure 4. Pro-HRnet-CNN convergence procedure

Optimizing HRNet with void-to-convolution boosts accuracy in identifying small nodules, preserving core benefits, as Figure 4 illustrates. Here, model iterations are on the X-axis, and test set detection performance on the Y-axis. Notably, accuracy surged initially over 6 epochs, gradually rose till the 21st epoch, achieving peak performance. After that, the precision changes tend to flatten out in the 22nd to 30th epoch, which indicates that the model has basically converged and entered a stable state at the 21st epoch.

Specific samples were selected from the test set, and various detection models were used to classify and locate lung images. The actual detection results of each model were visually presented in Figure 5, and the performance differences of different models in lung nodule recognition were compared and analyzed.

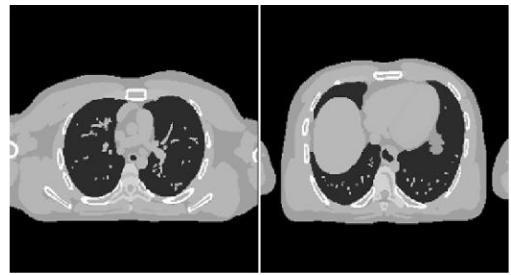

(a) Small nodules and microtubercles

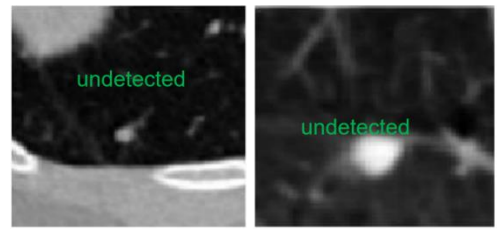

(b) Use ResNet-50 to test the effect

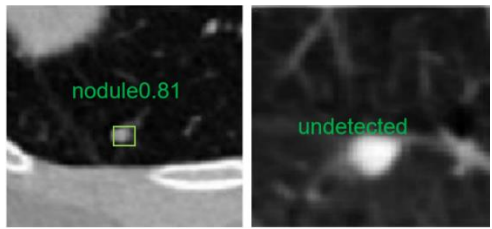

(c) The detection effect using the original HRNet

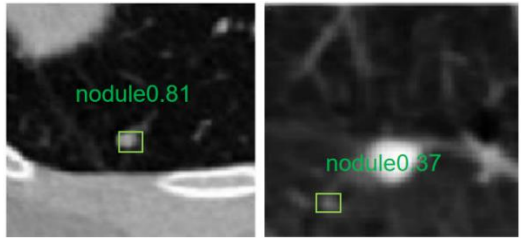

(d) Detection effect diagram of Pro-HRnet-CNN model

Figure 5. Comparison of lung CT imaging results under different backbone networks

Figure 5 depicts model predictions with green borders; "undetected" tags indicate missed samples. It showcases small and minimal nodules in 5(a), with 5(b)-5(d) illustrating varied models' performance on these.

Specifically, for small nodules, the Faster R-CNN model using ResNet-50 as the feature extractor has obvious missing phenomenon. On the contrary, using HRNet and Pro-HRNet as feature extraction layer models, these small nodules can be accurately detected even under high confidence conditions. Further, the very small nodules in Figure 5(a), even Faster R-CNN (based on ResNet-50) and original HRNET-driven models can not effectively identify, but by applying the Pro-HRnet-CNN model proposed in this chapter can successfully capture these subtle targets.

LIDC-IDRI dataset analysis confirms HRNet's superior small-nodule feature extraction over ResNet-50. Notably, Pro-HRNet-CNN enhances detection accuracy comparably to base HRNet, without extra parameters, showcasing potential in small target detection.

## V. CONCLUSIONS

This study demonstrates the innovative capabilities of the Pro-HRnet-CNN model, which combines HRNet with void convolution techniques to markedly enhance the detection of small pulmonary nodules in CT images. Our findings, supported by rigorous testing on the LIDC-IDRI dataset, clearly show that Pro-HRnet-CNN outperforms the traditional ResNet-50 model, particularly in detecting smaller nodules with higher accuracy. This enhancement is not only a technological advance but also a significant step toward more precise and early diagnosis of lung diseases. The potential implications of the Pro-HRnet-CNN model extend far beyond the realm of lung disease. Its application could revolutionize the fields of oncology, neurology, and cardiology, where early detection of small anomalies can drastically alter prognosis and treatment outcomes. For instance, in oncology, the ability to accurately detect small tumors at an early stage could lead to more effective intervention strategies, potentially increasing survival rates. Similarly, in cardiology, early detection of minute changes in heart tissue could aid in preventing severe cardiac events. Additionally, the integration of this model into real-time imaging systems in clinical settings could provide immediate and accurate diagnostic support, thus facilitating faster decision-making in emergency and critical care scenarios. This could significantly reduce diagnostic errors and improve patient outcomes. By advancing this technology, we move closer to a healthcare paradigm where precision medicine is not just a concept but a practical reality, enhancing patient care across multiple specialties and making early, accurate diagnosis accessible to all.

REFERENCES

[1] Boccatonda, A., Cocco, G., D'Ardes, D., Delli Pizzi, A., Vidili, G., De Molo, C., ... & Guagnano, M. T. (2023). Infectious pneumonia and lung ultrasound: a review. Journal of Clinical Medicine, 12(4), 1402.

[2] Yang, Y., Qiu, H., Gong, Y., Liu, X., Lin, X., & Li, M. (2024). Application of Computer Deep Learning Model in Diagnosis of Pulmonary Nodules. arXiv preprint arXiv:2406.13205.

[3] Gao, Z., Wang, Q., Mei, T., Cheng, X., Zi, Y., & Yang, H. (2024). An Enhanced Encoder-Decoder Network Architecture for Reducing Information Loss in Image Semantic Segmentation. arXiv preprint arXiv:2406.01605.

[4] Cheng, Y., Guo, J., Long, S., Wu, Y., Sun, M., & Zhang, R. (2024). Advanced Financial Fraud Detection Using GNN-CL Model. arXiv preprint arXiv:2407.06529.

[5] Xu, K., Wu, Y., Li, Z., Zhang, R., & Feng, Z. (2024). Investigating Financial Risk Behavior Prediction Using Deep Learning and Big Data. International Journal of Innovative Research in Engineering and Management, 11(3), 77-81.

[6] Xiao, M., Li, Y., Yan, X., Gao, M., & Wang, W. (2024, March). Convolutional neural network classification of cancer cytopathology images: taking breast cancer as an example. In Proceedings of the 2024 7th International Conference on Machine Vision and Applications (pp. 145-149).

[7] Zhan, Q., Sun, D., Gao, E., Ma, Y., Liang, Y., & Yang, H. (2024). Advancements in Feature Extraction Recognition of Medical Imaging Systems Through Deep Learning Technique. arXiv preprint arXiv:2406.18549.

[8] Hu, Y., Hu, J., Xu, T., Zhang, B., Yuan, J., & Deng, H. (2024). Research on Early Warning Model of Cardiovascular Disease Based on Computer Deep Learning. arXiv preprint arXiv:2406.08864.

[9] Feng, Y., Zhang, B., Xiao, L., Yang, Y., Gegen, T., & Chen, Z. (2024). Enhancing Medical Imaging with GANs Synthesizing Realistic Images from Limited Data. arXiv preprint arXiv:2406.18547.

[10] Xiao, L., Hu, J., Yang, Y., Feng, Y., Li, Z., & Chen, Z. (2024). Research on Feature Extraction Data Processing System For MRI of Brain Diseases Based on Computer Deep Learning. arXiv preprint arXiv:2406.16981.

[11] Zhu, Z., Yan, Y., Xu, R., Zi, Y., & Wang, J. (2022). Attention-Unet: A Deep Learning Approach for Fast and Accurate Segmentation in Medical Imaging. Journal of Computer Science and Software Applications, 2(4), 24-31.

[12] Liu, X., Qiu, H., Li, M., Yu, Z., Yang, Y., & Yan, Y. (2024). Application of Multimodal Fusion Deep Learning Model in Disease Recognition. arXiv preprint arXiv:2406.18546.

[13] Lassen BC, Jacobs C, Kuhnigk J.M, van Ginneken B. Robust semi-automatic segmentation of pulmonary subsolid nodules in chest computed tomography scans[J]. Physics in medicine and biology, 2015, 60: 1307-1323.

[14] Farahani FV et al. Lung nodule diagnosis from CT images based on ensemble learning[C]. IEEE, 2015.

[15] Javaid M et al. A novel approach to cad system for the detection of lung nodules in CT images[J]. Methods Programs Biomed, 2016, 135: 125-139.

[16] Roth HR et al. Improving Computer-Aided Detection using Convolutional Neural Networks and Random View Aggregation[J]. IEEE Trans. Med. Imaging, 2016, 35: 1170-1181.

[17] Yan, X., Wang, W., Xiao, M., Li, Y., & Gao, M. (2024, March). Survival prediction across diverse cancer types using neural networks. In Proceedings of the 2024 7th International Conference on Machine Vision and Applications (pp. 134-138).

[18] Setio AAA et al. Pulmonary nodule detection in ct images: False positive reduction using multi-view convolutional networks[J]. IEEE Trans. Med. Imaging, 2016, 35: 1160-1169.

[19] Liu, H., Li, I., Liang, Y., Sun, D., Yang, Y., & Yang, H. (2024). Research on Deep Learning Model of Feature Extraction Based on Convolutional Neural Network. arXiv preprint arXiv:2406.08837.

[20] Lin, Y., Li, M., Zhu, Z., Feng, Y., Xiao, L., & Chen, Z. (2024). Research on Disease Prediction Model Construction Based on Computer AI Deep Learning Technology. arXiv preprint arXiv:2406.16982.

[21] Hu, Y., Yang, H., Xu, T., He, S., Yuan, J., & Deng, H. (2024). Exploration of Multi-Scale Image Fusion Systems in Intelligent Medical Image Analysis. arXiv preprint arXiv:2406.18548.

[22] Sun, D., Liang, Y., Yang, Y., Ma, Y., Zhan, Q., & Gao, E. (2024). Research on Optimization of Natural Language Processing Model Based on Multimodal Deep Learning. arXiv preprint arXiv:2406.08838.

[23] Xiao, L., Li, M., Feng, Y., Wang, M., Zhu, Z., & Chen, Z. (2024). Exploration of Attention Mechanism-Enhanced Deep Learning Models in the Mining of Medical Textual Data. arXiv preprint arXiv:2406.00016.

[24] Skourt B A, El Hassani A, Majda A. Lung CT image segmentation using deep neural networks[J]. Procedia Computer Science, 2018, 127: 109-113.

[25] Sun, M., Feng, Z., Li, Z., Gu, W., & Gu, X. (2024). Enhancing Financial Risk Management through LSTM and Extreme Value Theory: A High-Frequency Trading Volume Approach. Journal of Computer Technology and Software, 3(3).

[26] Wang, J., Hong, S., Dong, Y., Li, Z., & Hu, J. (2024). Predicting Stock Market Trends Using LSTM Networks: Overcoming RNN Limitations for Improved Financial Forecasting. Journal of Computer Science and Software Applications, 4(3), 1-7.

[27] Xu, R., Zi, Y., Dai, L., Yu, H., & Zhu, M. (2024). Advancing Medical Diagnostics with Deep Learning and Data Preprocessing. International Journal of Innovative Research in Computer Science & Technology, 12(3), 143-147.